\documentclass[apjl,a4paper,12pt,useAMS]{emulateapj}
\usepackage{amsmath}
\usepackage{cases}
\usepackage{amssymb}
\usepackage{graphicx}
\usepackage{natbib}

\begin{document}

\title{Gravitational-wave memory from a propagating relativistic jet: a probe of the interior of gamma-ray burst progenitors}
\author{Yun-Wei~Yu$^{1,2}$}

\altaffiltext{1}{Institute of Astrophysics, Central China Normal
University, Wuhan 430079, China, {yuyw@mail.ccnu.edu.cn}}
\altaffiltext{2}{Key Laboratory of Quark and Lepton Physics (Central
China Normal University), Ministry of Education, Wuhan 430079,
China}

\begin{abstract}
It is believed that the relativistic jets of gamma-ray bursts (GRBs) should initially propagate through a heavy envelope of the massive progenitor stars or through a merger ejecta formed from the compact binary mergers. The interaction of a jet with a stellar envelope or a merger ejecta can lead to the deceleration of the head material of the jet and simultaneously the formation of a hot cocoon. However, this jet-envelope/ejecta interaction is actually undetectable with electromagnetic radiation and can only be inferred indirectly by the structure of the breakout jet. Therefore, as a solution to this phenomenon, we suggest that the jet-envelope/ejecta interaction can produce a gravitational-wave (GW) memory of an amplitude of $h\sim10^{-26}-10^{-23}$, which could be detected with some future GW detectors sensitive in the frequency range from sub-Hertz to several tens of Hertz. This provides a potential direct way to probe the jet propagation and then the interior of the GRB progenitors. Moreover, this method is in principle available even if the jet is finally chocked in the stellar envelope or the merger ejecta.
\end{abstract}

\section{Introduction}
Relativistic jets can be driven by the central engines of gamma-ray bursts (GRBs). Such jets should cross through a dense medium and be collimated there before generating the GRB emission. Specifically, for long-duration GRBs originating from the core collapses of massive stars \citep{Woosley1993,Woosley2006,Galama1998,Stanek2003,Hjorth2003}, the medium involved is just the heavy envelopes of the massive progenitors. For short-duration GRBs created by the mergers of compact binaries \citep{Paczynski1986,Eichler1989, Narayan1992,Abbott2017}, the dense medium surrounding the merger products can be contributed by various material ejections during the mergers (see \cite{Shibata2019} for a review). The interaction between the GRB jet and the envelope/ejecta can in principle provide plenty of information about the density profile of the envelope/ejecta. The uncovering of the interior of GRB progenitors is one of the primary purposes of GRB research. Unfortunately, this jet-envelope/ejecta interaction is in fact unobservable only with electromagnetic waves, because the envelope/ejecta is very opaque.

An indirect clue to the jet propagation could be found from the structures of the observable jet and cocoon that break out from the envelope/ejecta. Nevertheless, it is not easy to infer these structures by using the usual observations of GRB prompt and afterglow emissions. Recently, a very special opportunity had appeared in the joint observation of the gravitational-wave (GW) event GW170817 and the short GRB 170817A \citep{Abbott2017,Goldstein2017}. On the one hand, the GW detection robustly indicated that the line of sight of this event significantly deviates from the normal direction of the orbital plane of the merging binaries and thus deviates from the symmetric axis of the GRB outflow. On the other hand, tremendous interest has been successfully generated by the GW detection, leading to a very wide campaign of follow-up observations. As a result, the host galaxy, the associated kilonova, and the multi-wavelength afterglow emissions of GRB 170817A have been discovered subsequently and been monitored carefully and at length \citep{Coulter2017,Hallinan2017,Troja2017}. The prompt and afterglow emissions strongly suggest that the jet of GRB 170817A is highly structured. Its energy and velocity can decrease quickly with the increasing angle relative to the jet axis, which leads to the extremely low luminosity of GRB 170817A \citep{Goldstein2017,Zhang2018}. Therefore, in principle, the jet structure can be constrained by the dependence of the GRB rates on their luminosities, although which is very difficult because of the high uncertainty of the intrinsic luminosity function of GRB jets \citep{Gupte2018,Salafia2019,Tan2019}. More excitingly, for such an off-axis observed GRB, the angular distributions of the energy and Lorentz factor of the jet can even be directly derived from the afterglow light curves, because the lasting increase of the afterglow fluxes is just due to that the line of sight becoming closer and closer to the core of the jet \citep{Xiao2017,Mooley2018,Granot2018,Lyman2018,Lazzati2018,Ioka2018}.
As implied, the GRB jet is very likely to consist of a relativistic narrow core and a mildly relativistic wide-spreading wing. This jet structure undoubtedly offers a stringent constraint on the preceding interaction between the jet and the progenitor material.

Despite the encouraging implications of the jet structure, it is still very open-ended whether we can probe the jet-envelope/ejecta interaction more directly, in particular, in the rapidly developing era of GW astronomy. Recall that when a particle changes its velocity $v$ by interacting with other objects, the metric perturbation induced by this particle, $h\sim4GE\beta^2/(c^4 D)$, can change correspondingly, where $G$ is the gravitational constant, $E$ is the particle energy, $\beta=v/c$, $c$ is the speed of light, and $D$ is the luminosity distance of the particle to the observer. Consequently, a GW radiation can be generated as a response to this metric change \citep{Segalis2001}. Such a GW signal arising from an unidirectional metric variation (i.e., the metric perturbation does not return to its original value) is strictly not a periodic ``wave" but instead is usually called as a ``memory" \citep{Sago2004}. The characteristic frequency of a GW memory is determined by the timescale of the velocity change. Following this consideration, some authors have investigated the GW memory generated by radiating GRB jets \citep{Sago2004,Akiba2013,Birnholtz2013}.

Note that the propagation of a GRB jet is just a lasting process of (i) the continuous launching of new jet material from the central engine, (ii) the sharp deceleration of the jet material at its head, and (iii) the acceleration and expansion of the envelope/ejecta material. Therefore, a GW memory can be naturally expected to arise from these velocity jump processes. If detected, this GW memory can provide a direct probe of the jet-envelope/ejecta interaction, which can potentially help to determine the nature of the GRB progenitors. Tentatively, this paper is devoted to calculating the characteristic amplitude and frequency of the GW memory arising from the jet propagation, by using some reference density profiles for the envelope/ejecta material.

\section{The dynamics of jet propagation}
The dynamics of a relativistic jet propagating in an envelope/ejecta has been investigated widely by using both analytical and numerical methods \citep{Matzner2003,Bromberg2011,Levinson2013,Nakauchi2013,Nagakura2014,Bromberg2014,Bromberg2016,Matsumoto2018,Geng2019,Hamidani2019}. When a relativistic jet propagates into a medium, the collision between the jet and the medium can lead to a forward shock sweeping up the medium and a reverse shock accumulating jet material. The region between these two shocks can be called as the head of the jet. The hot material in the jet's head flows quickly and laterally to form a cocoon surrounding the jet. Then, the high pressure of the cocoon can drive a collimation shock in the jet material toward the jet axis. As a result, the jet can be gradually collimated until it breaks out from the medium.

Following the above considerations,
the velocity of a jet head can be determined by balancing the ram pressures of the forward-shocked envelope/ejecta and the reverse-shocked jet \citep{Matzner2003,Matsumoto2018}:
\begin{equation}\label{bh}
\beta_{\rm h}=\frac{\beta_{\rm j}\tilde L^{1/2}+\beta_{\rm e}}{\tilde L^{1/2}+1}
\end{equation}
with $\tilde L\simeq{L_{\rm j}}/({\Sigma_{\rm j}\rho_{\rm e}\Gamma_{\rm e}^2c^3})$,
where $\beta_{\rm j}\simeq1$, $L_{\rm j}$, and $\Sigma_{\rm j}$ are the velocity, one-sided luminosity, and cross section of the unshocked jet, respectively, and $\rho_{\rm e}$, $\beta_{\rm e}$, and $\Gamma_{\rm e}$ are the density, velocity, and Lorentz factor of the circum-material (i.e., the envelope/ejecta material), respectively.

The cross section of the jet can be calculated easily by
\begin{equation}\label{sigma1}
\Sigma_{\rm j}=\pi\theta_{\rm j0}^2 z_{\rm h}^2,
\end{equation}
if the jet collimation due to the cocoon pressure is weak and the jet keeps conical, where $\theta_{\rm j0}$ is the initial opening angle of the jet at launch and $z_{\rm h}=\int_0^{t}\beta_{\rm h}cdt'$ is the height of the jet head. Here the time $t$ is defined in the local rest frame. Nevertheless, in fact, the collimation effect is significant, once the jet height exceeds a critical point at $\hat{z}\approx (L_{\rm j}/\pi c P_{\rm c})^{1/2}$ where the collimation shock converges. The expression of $\hat{z}$ can be derived from the geometry equation of the collimation shock front \citep{Komissarov1997,Bromberg2011}, where $P_{\rm c}$ is the ram pressure in the cocoon. Due to the collimation, the jet cross section should be determined alternatively by \citep{Bromberg2011}
\begin{equation}\label{sigma2}
\Sigma_{\rm j}=\pi\theta_{\rm j0}^2 \left(\hat{z}\over 2\right)^2\simeq{\theta_{\rm j0}^2 L_{\rm j}\over 4cP_{\rm c}},
\end{equation}
where a height of $\hat{z}/2$ is used because at this point the jet shape starts to be transformed from conical to cylindrical. More specifically, at $\hat{z}/2$, the collimation shock becomes roughly parallel to the jet axis. In the following calculations, we use the smaller value by comparing Eqs. (\ref{sigma1}) and (\ref{sigma2}).

\begin{figure*}
\resizebox{0.45\hsize}{!}{\includegraphics{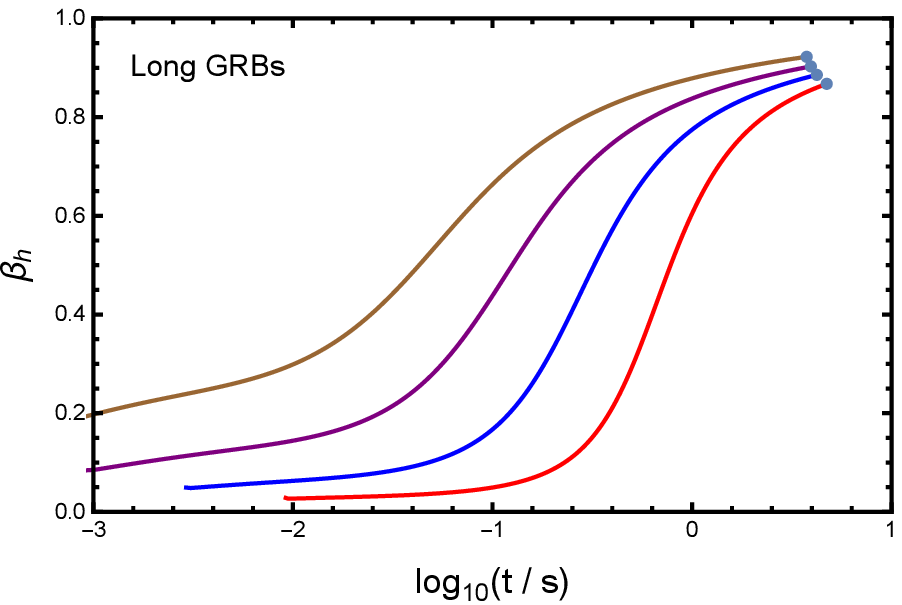}}\resizebox{0.45\hsize}{!}{\includegraphics{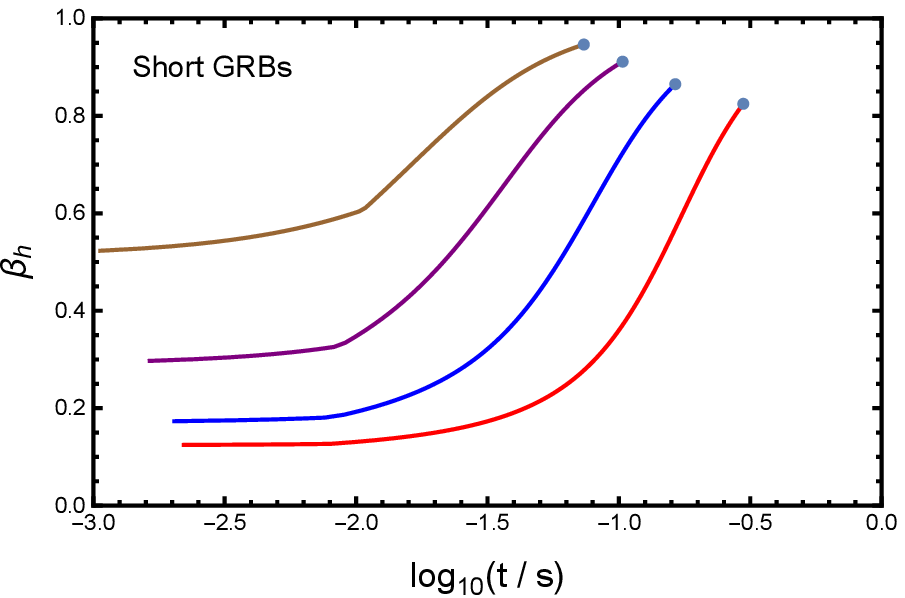}}
\caption{Evolution of the head velocity of a relativistic jet for different luminosities, i.e., $L_{\rm j}=10^{49}$, $10^{50}$, $10^{51}$, and $10^{52}\rm erg~s^{-1}$ from bottom to top. The other parameters are described in the text. The left and right panels correspond to long and short GRBs, respectively. The thick points represent the breakout of the jet. }\label{betah}
\end{figure*}

\begin{figure*}
\resizebox{0.45\hsize}{!}{\includegraphics{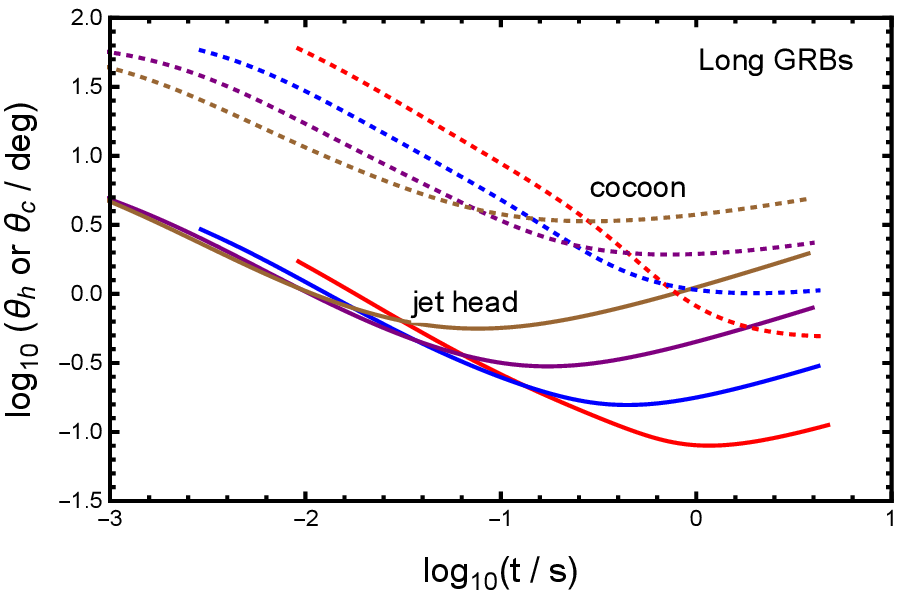}}\resizebox{0.45\hsize}{!}{\includegraphics{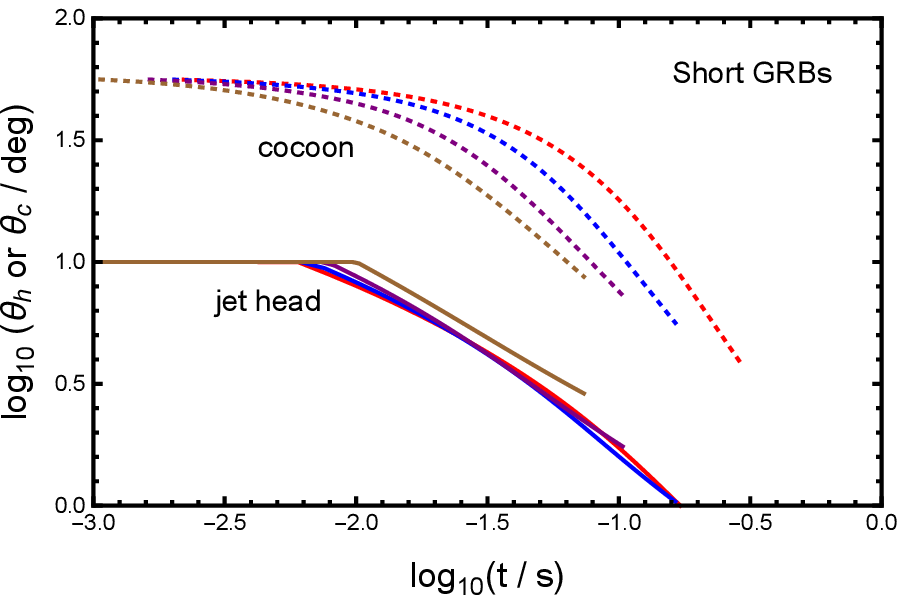}}
\caption{Evolution of the opening angles of the jet head (solid) and the cocoon (dashed) relative to the central engine.
}\label{thetat}
\end{figure*}

The pressure in the cocoon can be estimated by $P_{\rm c}={E_{\rm c}/( 3V_{\rm c}})$, by assuming it is radiation-dominated, where $E_{\rm c}$ is the total energy stored in the cocoon and $V_{\rm c}$ is the volume. By considering of the continuous energy injection from the jet head, the cocoon energy can be calculated by
\begin{equation}\label{Ec}
E_{\rm c}=\int_0^{t} L_{\rm j}\left(1-\beta_{\rm h}\right)dt'.
\end{equation}
The radius of the cross section of the cocoon is given by $r_{\rm c}=\int_0^{t}\beta_{\rm c,\perp}cdt'$, where the lateral expansion velocity reads $\beta_{\rm c,\perp} = [P_{\rm c}/\bar{\rho}_{\rm e}(z_{\rm h})c^2]^{1/2}$ \citep{Begelman1989} and $\bar{\rho}_{\rm e}=\int\rho_{\rm e}(z)dV/V_{\rm c}$ is the average density of the cocoon. The volume of the cocoon can be estimated by $V_{\rm c}=\pi r_{\rm c}^2z_{\rm h}$ by simply assuming a cylindrical shape.

The density profile of the circum-material can be usually described approximately by a power law. To be specific, for long GRBs, we use
\begin{equation}\label{rhozlgrb}
\rho_{\rm e}(z)=\rho_{\rm e,*}\left(1+{z\over z_*}\right)^{-\delta},{~\rm for~}z\leq z_{\rm s}
\end{equation}
with $\rho_{\rm e,*}=2.4\times10^9\rm g~cm^{-3}$, $z_*=10^8$ cm, $\delta=3.5$, and $z_{\rm s}=10^{11}$ cm, which correspond to a Wolf-Rayet star of a mass of $15M_{\odot}$ \citep{Suwa2011}. Here the stellar envelope is considered to be static with $\beta_{\rm e}=0$, by ignoring the relatively slow supernova expansion. For short GRBs, according to numerical simulations of binary neutron star mergers \citep{Hotokezaka2013}, the density profile of the merger ejecta can be written as
\begin{eqnarray}\label{rhozsgrb}
\rho_{\rm e}(z)={(3-\delta)M_{\rm ej}\over 4\pi z_{\min}^3}\left[1-\left({z_{\max}\over z_{\min}}\right)^{3-\delta}\right]^{-1}\left({z\over z_{\min}}\right)^{-\delta},\nonumber\\
{~\rm for~}z_{\min}\leq z\leq z_{\max},
\end{eqnarray}
where $M_{\rm ej}=0.01M_{\odot}$ is the total mass of the ejecta, $\delta=3.5$, and $z_{\max}=v_{\rm e,\max}t$ and $z_{\min}=v_{\rm e,\min}t$ are given for $v_{\rm e,\max}=0.4c$ and $v_{\rm e,\min}=0.1c$ \citep{Nagakura2014}.

As addressed, the dynamical evolution of the jet propagation is primarily dependent on the numerical ratio between the energy flux of the jet and the density of the circum-material. Therefore, in our calculations, we only take some reference density profiles and fix the initial opening angle to be $\theta_{\rm j0}=10^\circ$, while the value of $L_{\rm j}$ is varied within a wide range of $10^{49}-10^{52}\rm erg~s^{-1}$. Additionally, the luminosity is considered to not evolve with time. Then, we plot the velocity evolution of the jet head in Figure \ref{betah}, for different jet luminosities. Meanwhile, we also plot in Figure \ref{thetat} the evolution of the opening angles of the jet head and the cocoon relative to the central engine, which are defined as
\begin{eqnarray}\label{thetah}
\theta_{\rm h}=\left({\Sigma_{\rm j}\over \pi z_{\rm h}^2}\right)^{1/2},
\end{eqnarray}
and
\begin{eqnarray}\label{thetac}
\theta_{\rm c}={r_{\rm c}\over z_{\rm h}},
\end{eqnarray}
respectively. These angle variations reflect the degree of the jet collimation. It is shown that the edge of the jet is squeezed by the cocoon significantly and thus the jet material is pushed and accelerated. For both long and short GRBs, the higher the jet luminosity, the higher the head velocity and the shorter the time of the jet breakout $t_{\rm bo}$.

\section{The GW memory from a propagating jet}
The metric perturbation $h_{\mu \nu} = g_{\mu \nu}-\eta_{\mu \nu}$ induced by a point mass $m$ of a four-velocity $u^{\mu}(\tau)=\gamma c(1,\vec{\beta})$ can be solved from the following linearized Einstein equation (under the Lorentz gauge condition) \citep{Segalis2001}:
\begin{equation}
\left( -{1\over c^2}\frac{\partial^2}{\partial t^2}
+ \triangle \right) \bar{h}_{\mu \nu} = -16 \pi GT_{\mu \nu}\label{Einstein}
\end{equation}
with the energy-momentum tensor
\begin{equation}
T^{\mu \nu} (x) = \int m u^{\mu}(\tau) u^{\nu}(\tau)
\delta^{(4)}[x-x(\tau)] d \tau,
\end{equation}
where $\bar{h}_{\mu \nu} \equiv h_{\mu \nu}-\frac{1}{2} \eta_{\mu \nu} h^{\lambda}_{\lambda}$, $\eta_{\mu\nu}=(-1,+1,+1,+1)$ is the Minkowski metric, and $\tau$ is the proper time. By setting the $z-$axis at the velocity direction and transforming $h_{\mu\nu}$ from the Lorentz gauge to the  transverse-traceless (TT) gauge, the TT parts of $h_{ij}$ read \citep{Segalis2001,Sago2004,Akiba2013}
\begin{eqnarray}
h_+ \equiv h_{xx}^{\rm TT} &=& - h_{yy}^{\rm TT} = \frac{2 GE }{c^4D} \frac{\beta^2\sin^2 \vartheta}
{1-\beta \cos \vartheta} \cos 2 \varphi,  \label{hp}\\
h_{\times} \equiv h_{xy}^{\rm TT} &=& h_{yx}^{\rm TT} = \frac{2 GE }{c^4D} \frac{\beta^2\sin^2 \vartheta}
{1-\beta \cos \vartheta} \sin 2 \varphi,\label{hm}
\end{eqnarray}
which are useful for calculating the response of a GW detector. Here a general expression of energy $E$ is used to replace the original term $\gamma mc^2$, in order to take into account the metric perturbation induced by a hot moving material, e.g., shocked material.

In each second, an amount of new energy $L_{\rm j}$ can be supplied to a relativistic GRB jet by a central engine. This new energy is of course initially injected into the base of the jet. However, in a quasi-static case, the net effect of the energy injection can well be reflected by the prolongation of the jet at its top. Therefore, we consider that this energy to be effectively confined in a solid angle of $\Omega_{\rm h}=2\pi(1-\cos\theta_{\rm h})$. Simultaneously, in a unit solid angle, an amount of energy $\dot{E}_{\rm c}/\Omega_{\rm h}$ can be transferred from the jet to the head region through the reverse shock. This energy is subsequently shared by the shocked jet material and the shocked medium together, both of which can finally spread into a solid angle of $\Omega_{\rm c}=2\pi(1-\cos\theta_{\rm c})$. Then, according to Equations (\ref{hp}) and (\ref{hm}), the change rate of the GW memory contributed by a differential jet and cocoon can be written as
\begin{equation}
{d\dot{h}(\vartheta)\over d\Omega}=\left\{
\begin{array}{ll}
{2 G \over c^4D}\left({L_{\rm j}-\dot{E}_{\rm c}\over \Omega_{\rm h}}\frac{\sin^2 \vartheta}
{1-\cos \vartheta}+{\dot{E}_{\rm c}\over \Omega_{\rm c}}\frac{\beta_{\rm h}^2\sin^2 \vartheta}
{1-\beta_{\rm h}\cos \vartheta}\right),{~\rm for~}\theta\leq\theta_{\rm h},\\
{2 G \dot{E}_{\rm c}\over c^4D\Omega_{\rm c}}\frac{\beta_{\rm c}^2\sin^2 \vartheta}
{1-\beta_{\rm c}\cos \vartheta},{~\rm for~}\theta_{\rm h}<\theta\leq\theta_{\rm c},
\end{array}\right.\label{kernel}
\end{equation}
where the spherical coordinates $(\theta,\phi)$ are defined by setting the $z-$axis at the symmetric axis of the jet. The net velocity of the cocoon, $\beta_{\rm c}$, is considered to be comparable to the head velocity and much higher than the cocoon's lateral expansion velocity, i.e., $\beta_{\rm c,\perp}\ll\beta_{\rm c}\sim\beta_{\rm h}$. Strictly speaking, $\beta_{\rm c}$ could be somewhat lower than $\beta_{\rm h}$ and even be reversed, because backflows could appear in the cocoon. In any case, as long as $\beta_{\rm c}$ is sub-relativistic and especially we have $\Omega_{\rm c}\gg\Omega_{\rm h}$, the contribution to the memory radiation from the cocoon could always be subordinate to that from the relativistic jet.

The temporally accumulated amplitude of the GW memory can be calculated by the following integral:
\begin{eqnarray}
 h(t)=\sqrt{\left[\int_0^t\dot{h}_{+}(t')dt'\right]^2+ \left[\int_0^t\dot{h}_{\times}(t')dt'\right]^2}.\label{htobs}
\end{eqnarray}
with
\begin{eqnarray}
\dot{h}_+ (t)&=& \int_0^{2\pi}\int_0^{\theta_{\rm c}}{d\dot{h}(\vartheta)\over d\Omega}\cos 2 \varphi\sin\theta d\theta d\phi,  \label{Deltahp}\\
\dot{h}_{\times} (t)&=& \int_0^{2\pi}\int_0^{\theta_{\rm c}}{d\dot{h}(\vartheta)\over d\Omega}\sin 2 \varphi\sin\theta d\theta d\phi.  \label{Deltaht}
\end{eqnarray}
For a viewing angle $\theta_{\rm v}$ relative to the jet axis, the relationship between $(\vartheta,\varphi)$ and $(\theta,\phi)$ can be written as \citep{Akiba2013}
\begin{eqnarray}
\sin \vartheta \cos \varphi &=&
-\sin \theta_v \cos \theta + \cos \theta_v \sin \theta \cos \phi,  \\
\sin \vartheta \sin \varphi &=&
\sin \theta \sin \phi,  \\
\cos \vartheta &=& \cos \theta_v \cos \theta + \sin \theta_v \sin \theta \cos \phi.
\end{eqnarray}
As long as the viewing angle $\theta_{\rm v}\gg\theta_{\rm h}$, it can be found that the integrated memory amplitude as a function of $t$ can be simply approximated by
\begin{eqnarray}
h (t) \approx {2 G L_{\rm j}\beta_{\rm h}t\over c^4D}\frac{\sin^2 \theta_{\rm v}}
{1-\cos \theta_{\rm v}}\propto\beta_{\rm h}t,\label{htapp}
\end{eqnarray}
if the viewing angle $\theta_{\rm v}$ is not too small. This clearly indicates that the increase of the memory amplitude can basically trace the dynamical evolution of the jet propagation. Therefore, the detection of the GW memory can help us to shed light into the jet propagation occurring in GRB progenitors. It is worth mentioning that, in the frame of observers, the interval between the arriving times of GW memory signals should be defined by
\begin{eqnarray}
dt_{\rm obs}=dt\left(1-{\beta_{\rm h}\cos\theta_{\rm v}}\right).\label{tobs}
\end{eqnarray}
This treatment is the same as that adopted for GRB afterglow calculations. On the contrary, for GRB prompt emission and its consequent GW memory, the observer's time is equivalent to the local time, since the internal dissipation radius of a GRB jet is considered to be nearly stationary. In principle, the metric perturbation can also arise from the encounter jet, which can somewhat eliminate the contribution from the oncoming jet considered above. However, according to Equation (\ref{htapp}), the memory amplitude caused by the encounter jet should always be suppressed by a factor of about $(1-\cos\theta_{\rm v})/(1+\cos\theta_{\rm v})$ in comparison to the oncoming jet. Therefore, the contribution from the encounter jet can be roughly neglected.

\begin{figure}
\resizebox{0.9\hsize}{!}{\includegraphics{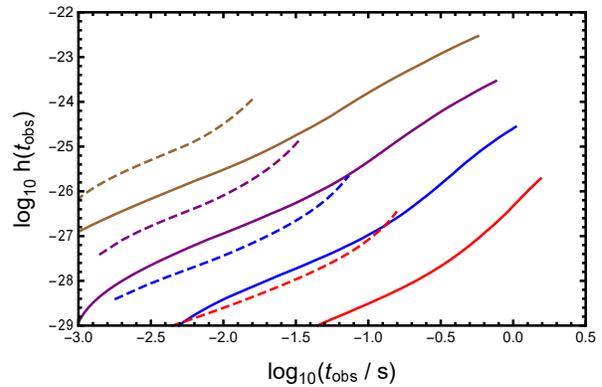}}
\caption{Increasing amplitude of the GW memory due to a jet propagation.
The solid and dashed lines correspond to long and short GRBs, respectively. A viewing angle of $\theta_{\rm v}=15^\circ$ is taken. The distances of the short and long GRBs are taken as 50 and 130 Mpc, respectively.}\label{ht}
\end{figure}

For a viewing angle of $\theta_{\rm v}=15^\circ$, in Figure \ref{ht} we present the increasing amplitudes of the GW memory of a propagating jet for different luminosities, where the distances of $D\sim130$ Mpc and $\sim50$ Mpc are taken for long and short GRBs, respectively. These reference distances are adopted by requiring that at least one GRB could happen in one year within these distances. In the near universe where the memory can be detected, we can roughly take the local event rates of long and short GRBs as $\sim100\rm ~Gpc^{-3}yr^{-1}$ \citep{Wanderman2010,Cao2011} and $\sim1500\rm ~Gpc^{-3}yr^{-1}$ \citep{Abbott2017}, respectively. These event rates are not discounted by the beaming effect, because the GW memory radiation is anti-beamed \citep{Sago2004,Akiba2013,Birnholtz2013} and cannot be obstructed by the circum-medium. On the contrary, the electromagnetic emission from a relativistic jet should be highly beamed within an angle of $\theta\sim\Gamma_{\rm j}^{-1}$. Therefore, the detection probability of the GW memory can in principle be much higher than that of its associated GRB emission. More importantly, the GW memory can always be detectable even though the jet is completely choked by the circum-medium. Note that the adopted short GRB rate has already included nearly all contributions from the mergers of neutron stars, whereas the long GRB rate only takes into account cases with a successfully jet breakout jet. Therefore, by considering of the potentially larger number of choked long GRBs, the reference distance for them could be much smaller. This is obviously beneficial for detecting GW memory from long GRBs.

\begin{figure}
\resizebox{0.9\hsize}{!}{\includegraphics{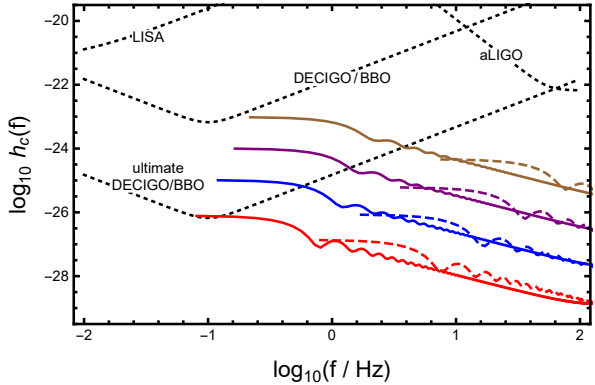}}
\caption{Characteristic amplitude of the GW memory as a function of the GW frequency. The solid and dashed lines correspond to long and short GRBs, respectively. The dotted lines give the noise amplitude of the GW detectors as labeled.}\label{hf}
\end{figure}

For a comparison with the noise amplitude of some GW detectors, we calculate the characteristic amplitude of the GW memory by
\begin{eqnarray}
h_{\rm c}(f)=2f|\tilde{h}|\equiv2f\left|\int_{-\infty}^{\infty}h(t_{\rm obs})e^{i2\pi ft_{\rm obs}}dt_{\rm obs}\right|,
\end{eqnarray}
where $f$ is the GW frequency. The obtained results are presented in Figure \ref{hf}. Here, limited by the resolution of the Fast Fourier transform that we use, these amplitude lines could only provide the envelope of rapidly oscillating lines, where many oscillations could not be shown in detail. In any case, it is showed that the characteristic frequencies of the GW memories are primarily around sub-Hertz for long GRBs and several to serval tens of Hertz for short GRBs, which are directly determined by the timescales of the jet breakout. In Figure \ref{hf}, the noise amplitudes of the GW detectors are defined by $h_n(f)=[5fS_h(f)]^{1/2}$ as in \cite{Sago2004}. The expressions of the spectral density of the strain noise $S_h(f)$ for aLIGO, LISA, and DECIGO/BBO, as well as its ultimate value, are taken from \cite{Finn2000} and \cite{Seto2001}. In comparison, we find that the GW memories due to the jet propagation of long GRBs could be promising targets for DECIGO/BBO at its ultimate sensitivity, both in their characteristic amplitudes and frequencies. On the contrary, those from short GRBs could be difficult to be discovered with these GW detectors.

After it breaks out from the progenitor material at the local time of $t_{\rm bo}$, the relativistic jet can be decoupled with the cocoon and move freely at a high Lorentz factor of $\Gamma_{\rm j}$. Here, if the jet launching continues, the memory amplitude could finally increase to $\sim{2 G L_{\rm j}(\beta_{\rm h}t_{\rm bo}+\Delta T)/c^4D}$, where $\Delta T$ is the GRB prompt duration, which is considered to be not very different from $t_{\rm bo}$. In any case, according to Equation (\ref{tobs}), the memory radiation during this period would be detected within a very short timescale of $\Delta T/2\Gamma_{\rm j}^2$ and thus its characteristic frequency is expected to be as high as $\sim 2\Gamma_{\rm j}^2/\Delta T$. This can be easily discriminated from the signals due to the jet-medium interaction. Therefore, the slight growth of the memory amplitude after the jet breakout is neglected in the calculations of Figure \ref{hf}.

\section{Summary and discussions}
The GW memory from a relativistic GRB jet propagating into a dense circum-material is investigated, where the circum-medium specifically corresponds to a stellar envelope or a merger ejecta. The characteristic frequency and amplitude of the GW memory are found to be potentially detectable with the future GW detectors such as DECIGO/BBO at its ultimate sensitivity, in particular for long GRBs. The detection of these GW memories will provide a potentially practical method for probing the interiors of the GRB progenitors, because the characteristic features of these memories are completely determined by the energy release of the GRB engines and the propagation dynamics of the relativistic jet (see Equation \ref{htapp}).

In principle, a relativistic jet has many opportunities to be choked by the circum-material. Then, the GRB emission would be unexpected, other than a gamma-ray/X-ray flash possibly arising from a hot cocoon. So, the jet-progenitor interaction definitely cannot be inferred by constraining the jet structure by using GRB observations. However, the GW memory due to the jet propagation before the final quenching can still occur and be detectable, although the metric perturbation will finally return to the initial value as the cocoon material completely diffuses into the envelope/ejecta. Additionally, for sufficiently small distances at which the memory can be detected, a nearly isotropic supernova/kilonova emission can still be expected to accompany with the memory signal. Then, these optical transients can be used to effectively localize and calibrate the memory signal. At the same time, the GW memory can conversely help to identify the special origins of these optical transients, which are associated with a failed GRB.  Additionally, note that the propagating GRB jets can also be detected through high-energy neutrinos, which can be produced by the internal and collimation shocks in the jets \citep{Horiuchi2008,Kimura2018}.

In summary, the GW detection in the frequency range of $\sim0.1-100$ Hz will definitely be helpful for extending our knowledge of the GRB phenomena, in particular the interiors of their different progenitors. However, technically, how to distinguish a GW memory signal from the huge number of events received by the GW detectors is still an open question. Then, a plausible solution is to develop a data analysis method based on various multi-messenger signals, which makes the electromagnetic counterpart emissions indispensable.

\acknowledgements
The author thanks the referees for their valuable comments. This work is supported by the
National Natural Science Foundation of China (grant No.
11822302 and 11833003) and the Fundamental
Research Funds for the Central Universities (grant
No. CCNU18ZDPY06).

\end{document}